\RequirePackage{fix-cm}
\documentclass[11pt]{article}

\usepackage{graphicx}
\usepackage{tikzit}

\usepackage{listings}
\usepackage{color}

\definecolor{dkgreen}{rgb}{0,0.6,0}
\definecolor{gray}{rgb}{0.5,0.5,0.5}
\definecolor{mauve}{rgb}{0.58,0,0.82}

\usepackage{url}

\usepackage{breakurl}

\usepackage{bm}

\RequirePackage[OT1]{fontenc}
\RequirePackage{amsmath}
\RequirePackage[numbers]{natbib}

\RequirePackage[colorlinks,citecolor=blue,urlcolor=blue]{hyperref}

\usepackage{graphicx}
\usepackage{amsmath}
\usepackage{graphics}
\usepackage{epsfig}
\usepackage{amssymb}
\usepackage{color}
\usepackage{float}
\usepackage[section]{placeins}
\usepackage{subcaption,graphicx}

\usepackage[a4paper,bindingoffset=0.2in,%
            left=0.5in,right=0.5in,top=0.5in,bottom=0.8in,%
            footskip=.25in]{geometry}

\newcommand{\be}{\begin{eqnarray}}
\newcommand{\ee}{\end{eqnarray}}
\newcommand{\bea}{\begin{eqnarray*}}
\newcommand{\eea}{\end{eqnarray*}}


\begin{document}

\title{Comparison of different exit scenarios from the lock-down for COVID-19 epidemic in the UK and assessing uncertainty of the predictions
}

\author{ Anatoly Zhigljavsky\footnote{Cardiff University} , Roger Whitaker$^*$, Ivan Fesenko\footnote{University of Nottingham}, Kobi Kremnizer\footnote{University of Oxford},  Jack Noonan$^*$ 
}

\maketitle

\abstract{
We model further development of the COVID-19  epidemic in the UK given  the current data and assuming  different  scenarios of handling the epidemic.
In this research, we further extend the  stochastic model suggested in \cite{us} and incorporate in it all  available to us  knowledge about parameters characterising the behaviour of the virus and the illness induced by it.
The models we use are flexible, comprehensive, fast to run  and allow us to incorporate the following:
\begin{itemize}
  \item   time-dependent strategies of handling the epidemic;
  \item  spatial heterogeneity of the population and heterogeneity of development of  epidemic in different areas;
   \item  special characteristics of particular groups of people, especially people with specific medical pre-histories and elderly.
\end{itemize}
Standard epidemiological models such as SIR and many of its modifications are not flexible enough and hence are not precise enough 
in the studies  that requires the use of the features above. Decision-makers get serious  benefits  from  using better and more flexible models 
as they can  avoid of nuanced lock-downs, better plan the exit  strategy based on local population data, different stages of the epidemic in different areas,
making specific recommendations to specific groups of people; all this  resulting in a lesser impact on economy, 
improved forecasts of regional demand upon NHS allowing for intelligent resource allocation.

In this work, we investigate the sensitivity of the model to all its parameters while considering several realistic scenarios of what is likely to happen with the dynamics of the epidemic after the lock-down, which has started on March 23, is lifted. 
 \\

The main findings from this research are the following:
\begin{itemize}
  \item
  { very little gain, in terms of the projected hospital bed occupancy and expected numbers of death, of continuing the lock-down beyond April 13,
provided the isolation of older and vulnerable people continues and the public carries on some level  of isolation in the next 2-3 months, see Section~\ref{sec:t2};}

  \item in agreement with \cite{us}, isolation of the group of   vulnerable  people during the next 2-3 months  should be one of the main priorities, see Section~\ref{sec:c};
  \item it is of high importance that the whole population  carries on some level of isolation in the next 2-3 months, see 
  Section~\ref{R2};

  \item the timing of the current lock-down seems to be very sensible in areas like London where the epidemic has started to pick up by March 23; in such areas the second wave of epidemic is not expected, see figures in Sections \ref{key} and \ref{sens};
  \item the epidemic should almost completely finish in July, no global second wave should be expected, except areas where the first wave is almost absent, see Section~\ref{sec:x}.
\end{itemize}

}

\section{The  model}

\label{sec2}

The main differences between the model we use in this work and the model of \cite{us} are the following:
\begin{itemize}
  \item in the current model, we use the split into mild and severe cases of illness, see Figure~\ref{flow};
  \item as a result, we can now better estimate the expected numbers of hospital beds required and expected deaths at time $t$; however, we still do not take into account important factors of hospital bed availability and different heterogeneities, see \cite{us} and Sections~\ref{rate} and \ref{key};
      \item     we include assessments of the sensitivity of our model, using Julia programming language  \cite{bezanson2017julia}; most of the previous epidemiological models do not allow this kind of sensitivity assessment as they have far too many parameters and very rigid conditions about the probability distributions involved.

\end{itemize}

\subsection{ Variables and parameters}

\begin{itemize}
\item $t$ - time (in days)
\item $t_1$ - the lock-down time, $t_1={\rm March\; 23}$
\item $t_2$ - time when the lock-down finishes
  \item $N$ - population size 
  \item $G$ - sub-population (e.g. a group of people aged $70+$)
   \item $n$ -  size of group $G$ 
   \item $\alpha=n/N $ (in the case of the UK and $G$ consisting of people aged $70+$, $\alpha=0.132$)
   \item $r_G$ - average mortality rate in the group $G$
   \item $r_{other}$ - average mortality rate for the rest of population
  \item $I(t)$ - number of infectious at time $t$
  \item $S(t)$ - number of susceptible at time $t$;

  \item $1/\sigma_M$ - average time for recovery in mild cases
  \item $1/\sigma_S$ - average time until recovery (or death) in severe cases
    \item $\lambda_I$ - mean of the incubation period during which an infected person cannot spread the virus
\item $\delta$ - the probability of death in severe cases
   \item $R_0$  - reproductive number (average number of people who will capture the disease from one contagious  person)
         \item  $R_1$  - reproductive number  during the lock-down ($t_1\leq t<t_2$)
         \item  $R_2$  - reproductive number  after the lock-down finishes  ($t\geq t_2$)
        \item $c$ - the degree of separation of vulnerable people
        \item $x$ - the proportion of infected at the start of the lock-down
\end{itemize}

\subsection{ Values of parameters and generic model}

The  reproductive number $R_0$  is the main parameter defining the speed of development of an epidemic.
There is no true value for  $R_0$ as it varies  in different parts of the UK (and the world).
In particular, in rural areas one would expect a considerably lower value of $R_0$ than in London.
 Authors of \cite{Ferguson} suggest $R_0=2.2$ and $R_0=2.4 $ as typical; the authors of \cite{Lourenco} use values for $R_0$ in the range [2.25, 2.75].
We shall use the value $R_0=2.5$ as typical, which, in view of the recent data, looks to be  a rather high (pessimistic)  choice  overall. However,
$R_0=2.5$ seems to be  an adequate choice for the mega-cities where the epidemics develop faster and may lead to more causalities. In rural areas, in small towns, and everywhere else where social contacts are less intense,  the epidemic is milder.

\begin{figure}[h]
\begin{center}
\hspace{-3cm} 
\begin{tikzpicture}[scale=0.5]
        \begin{pgfonlayer}{nodelayer}
                \node [style=none] (6) at (-17, 0) {};
                \node [style=none] (7) at (-13, 0) {};
                \node [style=none] (8) at (-10, 3) {};
                \node [style=none] (9) at (-10, -3) {};
                \node [style=none] (10) at (-6, 3) {};
                \node [style=none] (11) at (-3, -3) {};
                \node [style=none] (12) at (0, -1) {};
                \node [style=none] (13) at (0, -5) {};
                \node [style=none] (14) at (-3.5, 3) {\footnotesize Alive};
                \node [style=none] (15) at (-15, -0.75) {};
                \node [style=none] (16) at (-15, -1.25) {$\tau_I$};
                \node [style=none] (17) at (-10.5, 1.25) {\footnotesize Mild};
                \node [style=none] (18) at (-10.25, -1.25) {\footnotesize Severe};
                \node [style=none] (19) at (-6.5, -4.25) {$\tau_S$};
                \node [style=none] (20) at (-8, 4.25) {$\tau_M$};
                \node [style=none] (21) at (-2, -1.5) {$1\!-\!\delta$};
                \node [style=none] (22) at (-2, -4.25) {$\delta$};
                \node [style=none] (24) at (1, -1) {\footnotesize Alive};
                \node [style=none] (25) at (1, -5) {\footnotesize Dead};
                \node [style=none] (26) at (-12.5, 2) {$1\!-\!q$};
                \node [style=none] (27) at (-12, -2) {};
                \node [style=none] (28) at (-12, -2) {$q$};
                \node [style=none] (29) at (-10.25, 3) {};
                \node [style=none] (30) at (-10.25, -3) {};
                \node [style=none] (31) at (-16.75, -0.25) {};
                \node [style=none] (32) at (-13.25, -0.25) {};
                \node [style=none] (33) at (-9.75, 3.25) {};
                \node [style=none] (34) at (-6.25, 3.25) {};
                \node [style=none] (35) at (-8, 3.75) {};
                \node [style=none] (36) at (-6.5, -3.75) {};
                \node [style=none] (37) at (-6.5, -3.75) {};
                \node [style=none] (38) at (-9.75, -3.25) {};
                \node [style=none] (39) at (-3.25, -3.25) {};
                \node [style=none] (40) at (-4.75, 3) {};
        \end{pgfonlayer}
        \begin{pgfonlayer}{edgelayer}
                \draw [|-|, style=none, in=180, out=0, looseness=0.75] (6.center) to (7.center);
                \draw [|-|, style=none] (8.center) to (10.center);
                \draw [|-|, style=none] (9.center) to (11.center);
                \draw [->] (11.center) to (12.center);
                \draw [->] (11.center) to (13.center);
                \draw [->] (7.center) to (29.center);
                \draw [->] (7.center) to (30.center);
                \draw [in=-75, out=105, looseness=0.75] (15.center) to (31.center);
                \draw [in=-90, out=90, looseness=0.75] (15.center) to (32.center);
                \draw [in=-90, out=90, looseness=0.50] (33.center) to (35.center);
                \draw [in=-90, out=90, looseness=0.50] (34.center) to (35.center);
                \draw [in=-90, out=90, looseness=0.25] (37.center) to (38.center);
                \draw [in=-90, out=90, looseness=0.25] (37.center) to (39.center);
                \draw [->] (10.center) to (40.center);
        \end{pgfonlayer}
\end{tikzpicture}
\end{center}
\caption{Flow-chart for the process of illness} \label{flow}
\end{figure}
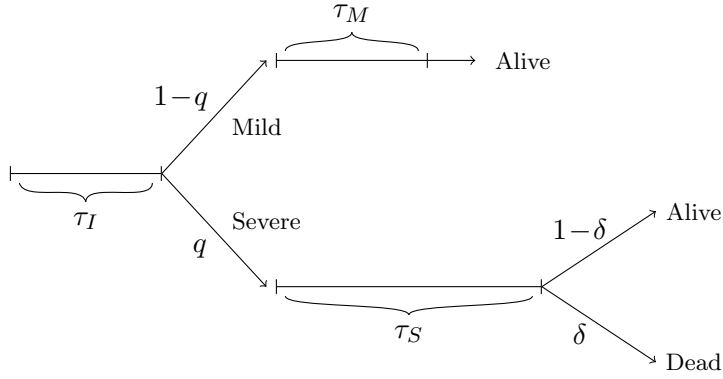

The flow-chart in Figure~\ref{flow} describes the process of illness. 
We assume that the person becomes infected $\tau_I$ days after catching the virus,
where $\tau_I$ has Poisson distribution with mean of $\lambda_I$ days. Parameter $\lambda_I$ defines the mean of the incubation period during which the person cannot spread the virus. Recommended value for $\lambda_I$ is $\lambda_I=5$.

{We then assign probabilities $q$ and $1-q$  for a person to get a severe or mild case correspondingly.
The value of $q$ depends on whether the person belongs to the group $G$ (then $q=q_G$) or the rest of population (in this case, $q=q_{other}$). In Section~\ref{rate} below we will  relate the probabilities $ q_{G}$ and $  q_{other}$
to the mortality rates in the two groups.

In a mild case, the person stays infectious for $\tau_M$ days and then discharges alive (that is, stops being infectious). We assume that the continuous version of $\tau_M$
has  { Erlang distribution} with shape parameter $k_M$ and rate parameter $\lambda_M$,  the mean of this distribution is $k/\lambda_M= 1/\sigma_M$ (in simulations, we discretise the numbers to their nearest integers). We use values $k_M=3$ and $\lambda_M=1/2\,$ so that $E\tau_M=k_M/\lambda_M =1/\sigma_M=6$. The variance of $\tau_M$ is
${\rm var}(\tau_M)=k_M/\lambda_M^2$; for  $k_M=3$ and $\lambda_M=1/2\,$ we have ${\rm var}(\tau_M)=12$ which seems to be a reasonable value.

In a  severe case, the person stays infectious for $\tau_S$ days. The continuous version of $\tau_S$
has  { Erlang distribution} with shape parameter $k_S$ and rate parameter $\lambda_S.$
The  mean of this distribution is $k_S/\lambda_S= 1/\sigma_S$. We use values $k_S=3$ and $\lambda_S=1/7\,$ so that $E\tau_S=k_S/\lambda_S =21$,
 in line  with the current knowledge, see e.g. \cite{Tang,Kucharski,womenshealth}. The variance of $\tau_S$ is
${\rm var}(\tau_S)=k_S/\lambda_S^2$; for  $k_M=3$ and $\lambda_S=1/7\,$ the standard deviation of the chosen Erlang distribution is approximately 12, which is rather large and reflects the uncertainty we currently have about the period of time a person needs to recover (or die) from COVID-19. 

In a severe case, the person dies with probability $\delta>0$ on discharge. The relation between $\delta$, the average mortality rates, the probabilities $q_S$ for the group $G$ and the rest of population is discussed in Section~\ref{rate}.

Note that the use of Erlang distribution is standard for modelling similar events in reliability and queuing theories, which have much in common with epidemiology. 

\subsection{Probability of getting a severe case,  probability of death in case of severe case $\delta$ and mortality rates}
\label{rate}

In this section, we relate the probability  of getting  severe case from group $G$ and the rest of population to the average mortality rates. 

We define $G$ as the group of vulnerable people. In the computations below, we assume that $G$ consists of people aged $70+$. 
We would like to emphasise, however, that there is still a lot of uncertainty on who is vulnerable. Moreover, it is very possible that the long term effect of the virus
might cause many extra morbidities in the coming years coming from severe cases who recover.

 We use the common split of the UK population into following age groups:
\bea
\;G_1=[0,19],\;\; G_2=[20,29],\; G_3=[30,39],\;
G_4=[40,49],\; \\G_5=[50,59], \;G_6=[60,69], \;G_7=[70,79], \;G_8=[80,\infty)
\eea
and corresponding  numbers $N_m$ ($m=1, \ldots, 8$; in millions) taken from \cite{statista}
$$[15.58, 8.71, 8.83, 8.50, 8.96, 7.07, 5.49, 3.27] \;\; {\rm with} \; N=66.41$$

The  probabilities of death for group $G_m$, denoted by $p_m$, are given from Table 1 in \cite{Ferguson} and replicated many times by the BBC and other news agencies are:
\bea
\label{eq:prob}
[ 0.00003, 0.0003, 0.0008, 0.0015, 0.006, 0.022, 0.051, 0.093]\, .
\eea
Unfortunately, these numbers do not match the other key number given in \cite{Ferguson}: the UK average mortality rate which is estimated to be  about $0.9\%.$
As we consider the value of the  UK average mortality rate as more important, we have multiplied all  probabilities above by
0.732 to get the average mortality rate to be  $0.9\%.$

Moreover, recent COVID-19 mortality data shows that the average mortality rates, especially  for younger age groups, are significantly lower than the ones given above. As we do not have reliable sources for the average mortality rates, we shall use the data from \cite{Ferguson} adjusted to 
the average mortality rate to be  $0.9\%$. As we shall be getting better estimates of the average mortality rates, the model can be easily adjusted to them.

There is an extra controversy which concerns the definition of what is death caused by COVID-19. As an example, the following is a direct quote from the official ONS document \cite{ons}: {\it "In Week 13, 18.8\% of all deaths mentioned “Influenza or Pneumonia”, COVID-19, or both. In comparison, for the five-year average, 19.6\% of deaths mentioned “Influenza and Pneumonia”. “Influenza and Pneumonia” has been included for comparison, as a well-understood cause of death involving respiratory infection that is likely to have somewhat similar risk factors to COVID-19."} In view of facts like this, the true COVID-19 mortality rates could be up to 5-10 times lower than given in \cite{Ferguson} and used in this work.

If consider population data for the whole UK and  define  the group $G$ as a union of groups $G_7$ and $G_8$, we have $n=5.49+3.27= 8. 76$m and $\alpha= 8.76/66.41 \simeq 0.132$.
The mortality rate in the group $G$ and for the rest of population  are therefore
$$
r_G= \frac{N_7 p_7+N_8 p_8}{n}\simeq 0.049; \;\; r_{other}= \frac{N_1 p_1+\ldots+ N_6 p_6}{N-n }\simeq  0.0030. \;\;
$$

Consider now a random person from the group $G$ who has got an infection. The probability he dies is $r_G$. On the other hand, this is also the product of the probability that this person has a severe case (which is $q_G$) and the probability (which is $\delta$) that the person dies conditionally he/she has severe case. The same is true for the rest of population. Therefore,
$$
\fbox{ \mbox {$q_G=r_G/\delta\;\;\; {\rm and } \;\;  q_{other}= r_{other}/\delta\, .$}}
$$

The COVID-19 epidemic  in the whole of the UK is subject to  different heterogeneities discussed in detail in \cite{us}. The model above assumes homogeneity and hence cannot be applied to the whole of the UK and we feel it would be irresponsible to estimate the total death toll for the UK using such a model. Instead, we shall assume that we apply this model to the population of inner London with population size rounded to 3 million. It may be tempting to multiply our expected death tolls by 22=66/3 but this would give very elevated forecasts. Indeed, epidemic in inner London can be considered
as the worst-case scenario and, in view of \cite{us}, we would recommend  to down-estimate the UK overall death toll forecasts by a factor of 2 or even more. 

\section{Main scenarios }

In this section we formulate two main scenarios and explain the main figures. In Section~\ref{sens}
we study  sensitivity with respect to all model parameters.

\subsection{Two  plots for the main set of parameters, no intervention}
\label{key}

As the main set of parameters for the main model we use the following: 
$$
\fbox{\mbox{$R_0=2.5$,  $\lambda_I=5$, $k_M=k_S=3$,    $\sigma_S=1/21$, $\sigma_M=1/5$, $\alpha=0.132$,  $\delta=0.2$.} }
$$
Note that in Section~\ref{key} we shall introduce four further parameters which will describe the lock-down of March 23 and the situation at the exit from this lock-down.

In Figure~\ref{fig1}  we plot, in the case with no intervention,  proportions of infected at time $t$ (either overall or in the corresponding group) and use the following colour scheme:
\begin{itemize}
  \item Solid green: $I(t)$, the proportion of infected in  overall population at time $t$.
  \item Purple: $I_G(t)$, the proportion of infected in group $G$ at time $t$
  \item Dashed green: the proportion of infected in the rest of population.
\end{itemize}
Knowing proportion of infected at time $t$ is important for knowing the danger 
of being infected for non-infected and non-immune  people.

\begin{figure}[h]
\centering
\begin{minipage}{.45\textwidth}
  \centering
  \includegraphics[width=1\textwidth]{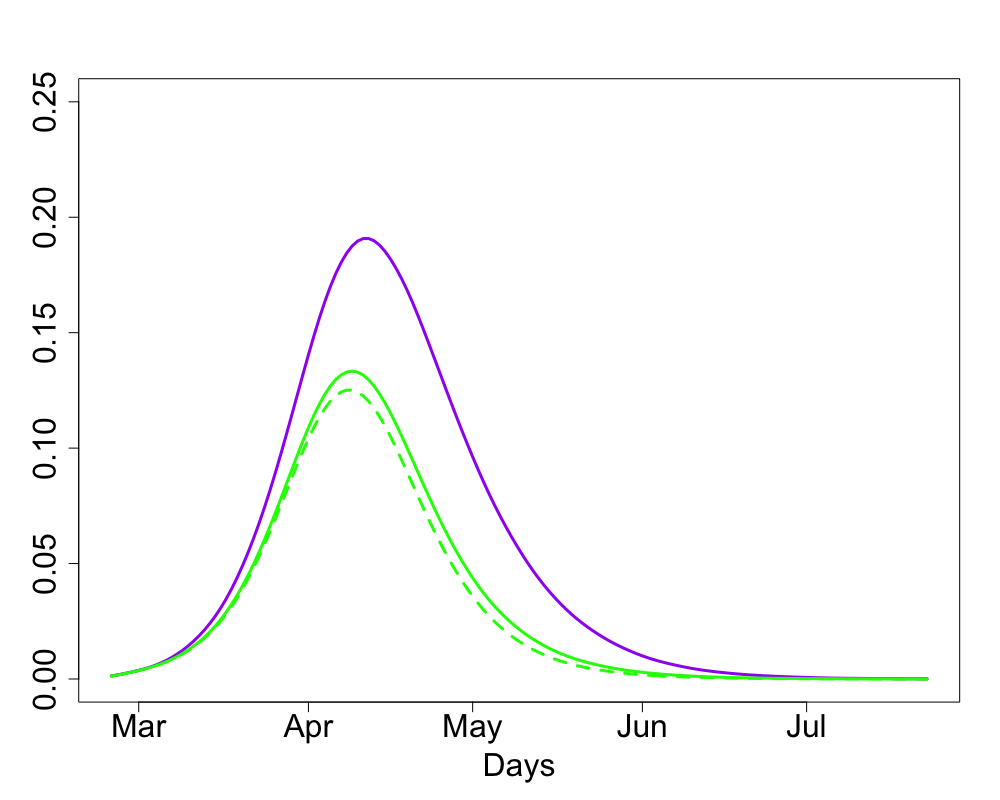}
\caption{Proportions of people infected \\ at time $t$ (overall, group $G$ and the  rest\\ of population).}
\label{fig1}
\end{minipage}%
\begin{minipage}{.45\textwidth}
  \centering
\includegraphics[width=1\textwidth]{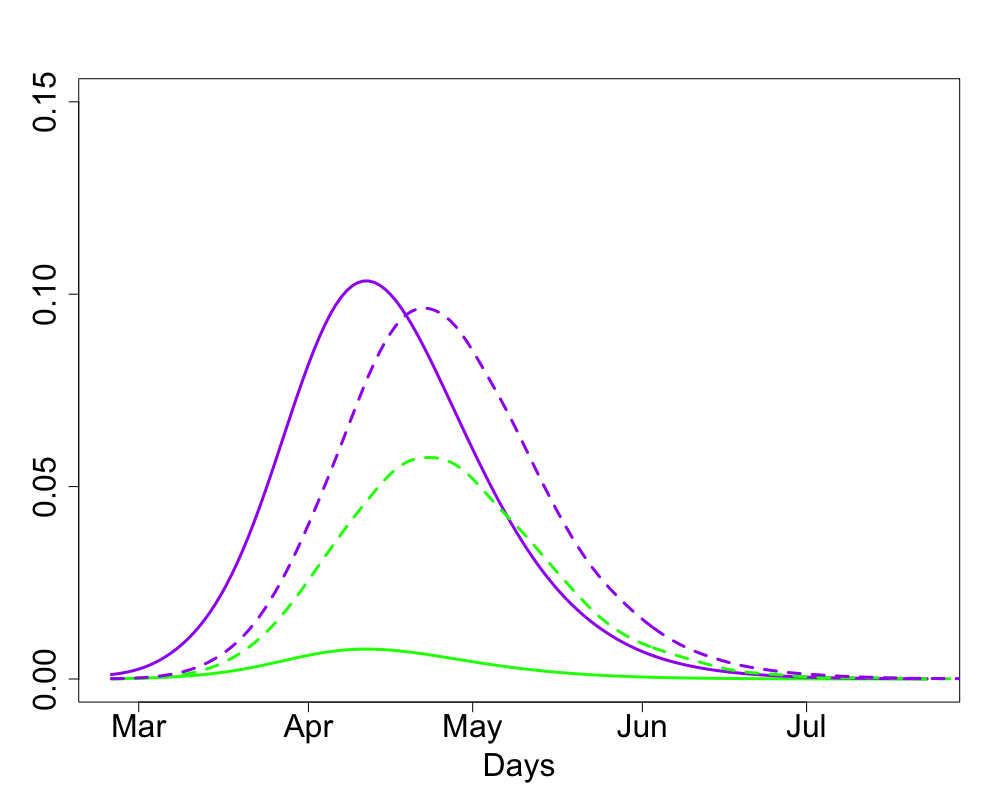}
\caption{Proportions of people with severe case of disease at time~$t$ (solid lines) and died at time $t$ (dashed lines). }
\label{fig2}
\end{minipage}
\end{figure}

\begin{figure}[ht]
\centering
\begin{minipage}{.45\textwidth}
  \centering
  \includegraphics[width=1\textwidth]{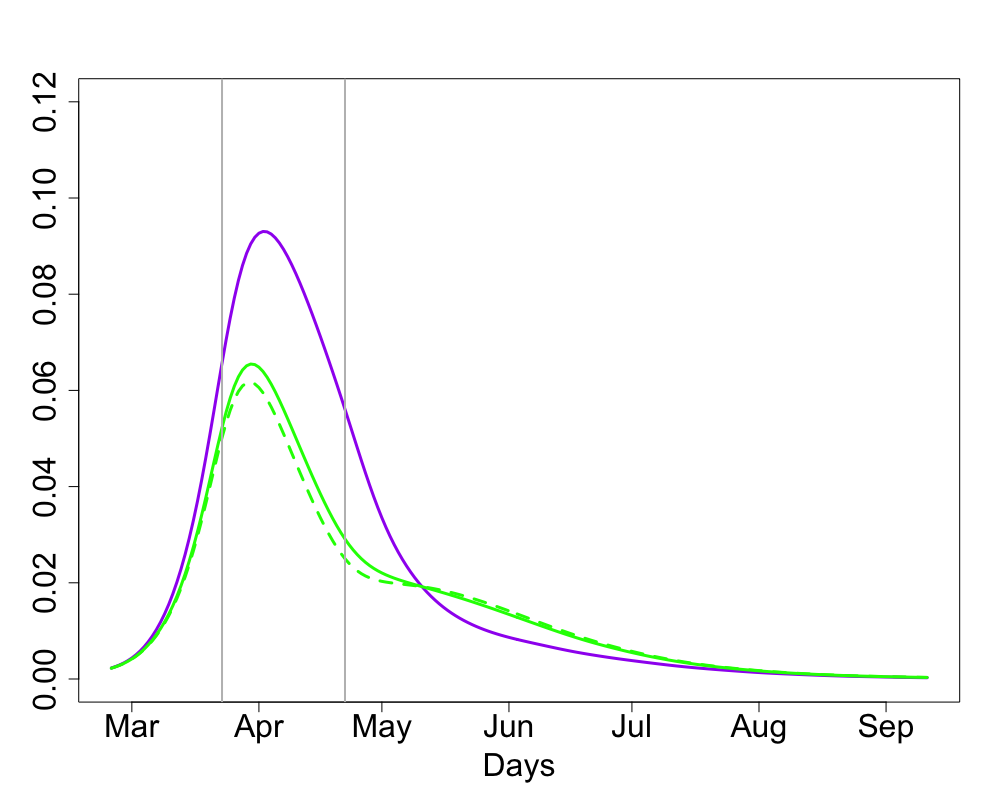}
\caption{Proportions of people infected \\ at time $t$ (overall, group $G$ and the  rest\\ of population).}
\label{fig3}
\end{minipage}%
\begin{minipage}{.45\textwidth}
  \centering
\includegraphics[width=1\textwidth]{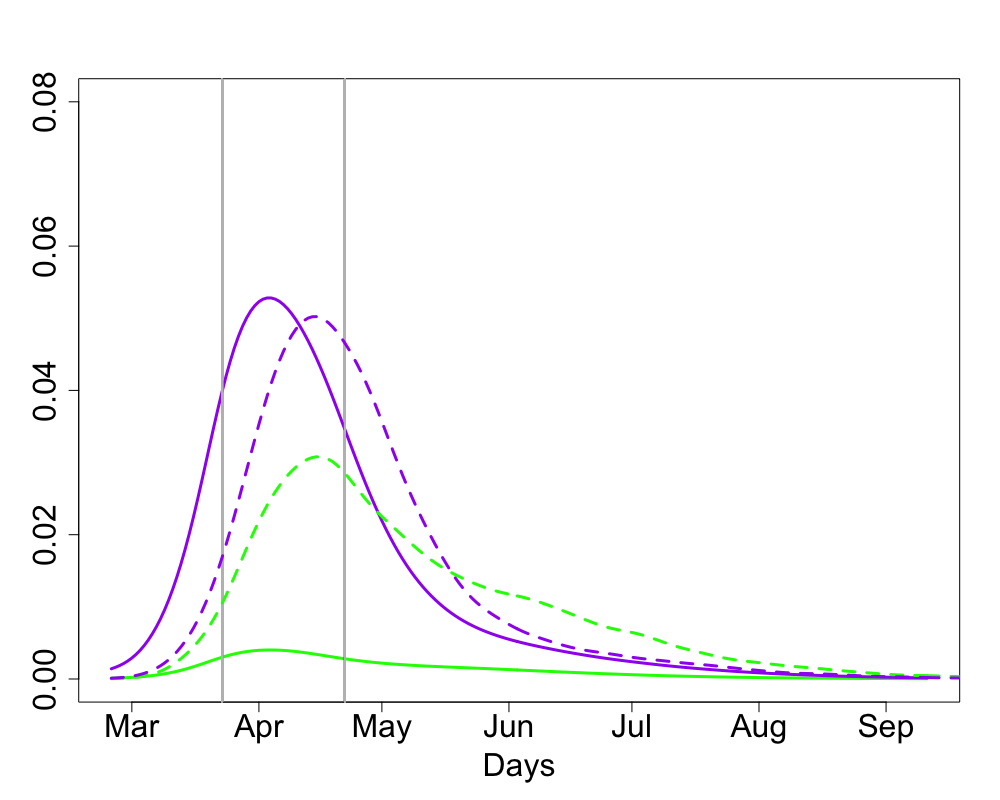}
\caption{Proportions of people with severe case of disease at time $t$ (solid lines) and discharged at time $t$ (dashed).
Expected death toll: 15K}
\label{fig4}
\end{minipage}
\end{figure}

In Figure~\ref{fig2} we plot, in the case of no intervention,  proportions of people with severe case of disease at time $t$ and proportions of such people discharged
 at time $t$; there are separate plots for the group $G$ and the rest of population.

\begin{itemize}
  \item Solid purple: proportion of people from group $G$ with severe case at time $t$.
  \item Solid green: proportion of people from the rest of population with severe case at time $t$.
   \item Dashed purple: proportion of people  from group $G$ died at time $t$, multiplied by 100.
  \item Dashed green: proportion of people  from the rest of population died at time $t$, multiplied by 1000 (so that the curve can be seen in the figure).
\end{itemize}
Proportions of people  with severe case at time $t$ is the main characteristic needed for planning NHS work-load. 
The proportion of people with severe case from group $G$ discharged at time $t$ is proportional to the expected number of deaths after multiplying this number by $\delta$ and  the size of the group $G$; similar calculations can be done for the rest of population.
Note that in these calculations we do not take into consideration an extremely important factor of  hospital bed availability  are at time $t$. Currently, we do not have a model for this.

We provide the values of the overall expected death toll. In the case of no lock-down, this toll would be 24(7+17)K, where the first number in the bracket correspond to the rest of population and the second one to the group $G$.

\subsection{Key plots for the main set of parameters, the case with intervention}
\label{key}

In this section, we study the scenario where we have made an intervention on $t_1={\rm March\; 23}$ by reducing $R_0$ to $R_1<R_0$ and then terminating the lock-down on $t_2={\rm April\; 22}$.

As we do not know at which stage of the epidemic the lock-down has started,  we set  a parameter to denote it. This parameter is a number $x \in (0,1)$ such that $S(t_1)/N=x$.

After lifting the lock-down, we   assume  that people from the group $G$ are still isolated but the rest of population returns back to normal life. However, it is natural to assume that the reproduction number $R_2$ for $t \geq t_2$ is smaller that the initial value of $R_0$. To denote the degree of isolation of people from $G$ for $t \geq t_2$,
we assume that for $t\geq t_2$ the virus is  transmitted to people in such a way that, conditionally a virus is transmitted, the probability that it reaches  a person from $G$ is $p=c \alpha$ with $0<c\leq 1$. Our main value for $c$ is $1/4$.
This means  that for $t\geq t_2$, under the condition that a virus is  infecting a new person,  the probabilities  that this new person belongs to $G$ is $1/5$.

Measuring the level of compliance in the population and converting this to simple epidemiological measures $c$, $ R_1$ and $ R_2$ and is hugely complex
problem which is beyond the scope of this paper.

Summarizing, there are five  parameters modelling the lock-down: 
$$
\fbox{\mbox{ $t_2={\rm April\; 22}$, $x=0.9$,
$R_1=1$,  $R_2=2$, $c=1/4$.
}}
$$

Figures~\ref{fig3} and~\ref{fig4} similar to  Figures~\ref{fig1} and~\ref{fig2} but computed for the case of intervention.
The colour schemes in Figures~\ref{fig3} and~\ref{fig4} are exactly the same as in  Figures~\ref{fig1} and~\ref{fig2}, respectively. 
The two grey lines in Figures~\ref{fig3} and~\ref{fig4} mark the intervention times. Note different scales  in the $y$-axis  in Figures~\ref{fig3} and ~\ref{fig4} in comparison to Figures~\ref{fig1} and~\ref{fig2}

We will study sensitivity of the curves $I(t)$ and $I_G(t)$, plotted in Figures~\ref{fig3} and again in Figure~\ref{fig5} for the same parameters, with respect to all parameters of the model. In Figure~\ref{fig6} we plot average death numbers (purple for group $G$ and green for the rest of the population) in the case of the inner London as  explained at the end of Section~\ref{rate}. Note that these numbers do not take into account the important factor of hospital bed availability and various heterogeneities, as explained in \cite{us}.

\begin{figure}[h]
\centering
\begin{minipage}{.45\textwidth}
  \centering
  \includegraphics[width=1\textwidth]{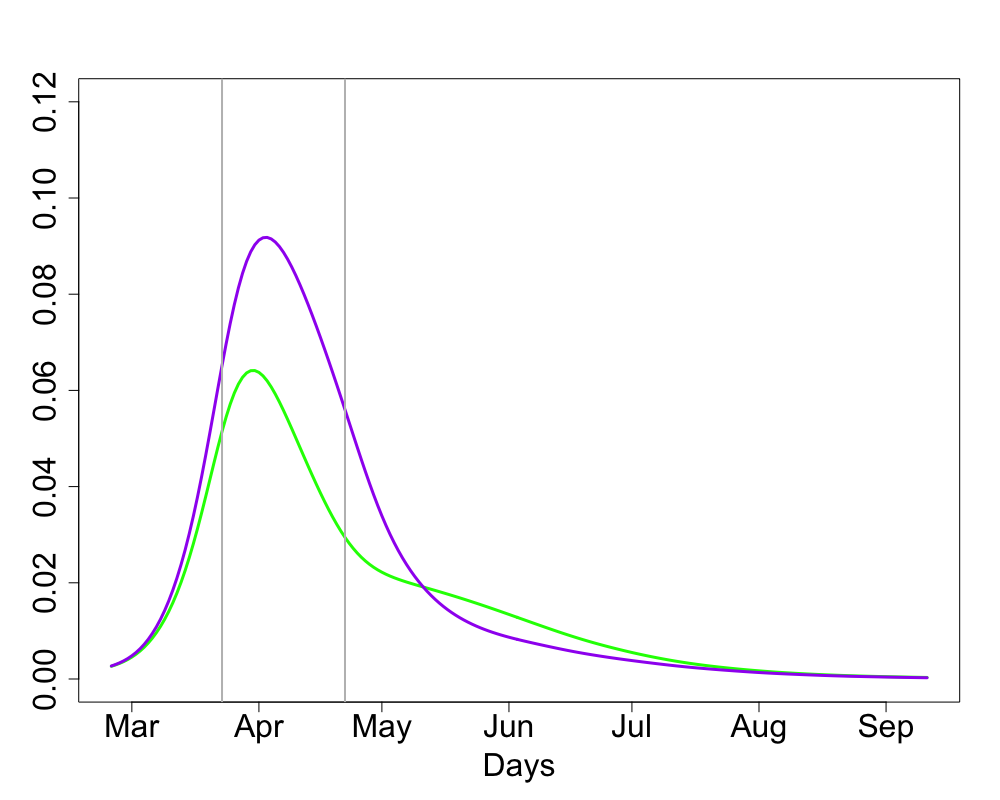}
\caption{Proportions of people infected \\ at time $t$ (group $G$ and overall).\\
$\,$}
\label{fig5}
\end{minipage}%
\begin{minipage}{.45\textwidth}
  \centering
\includegraphics[width=1\textwidth]{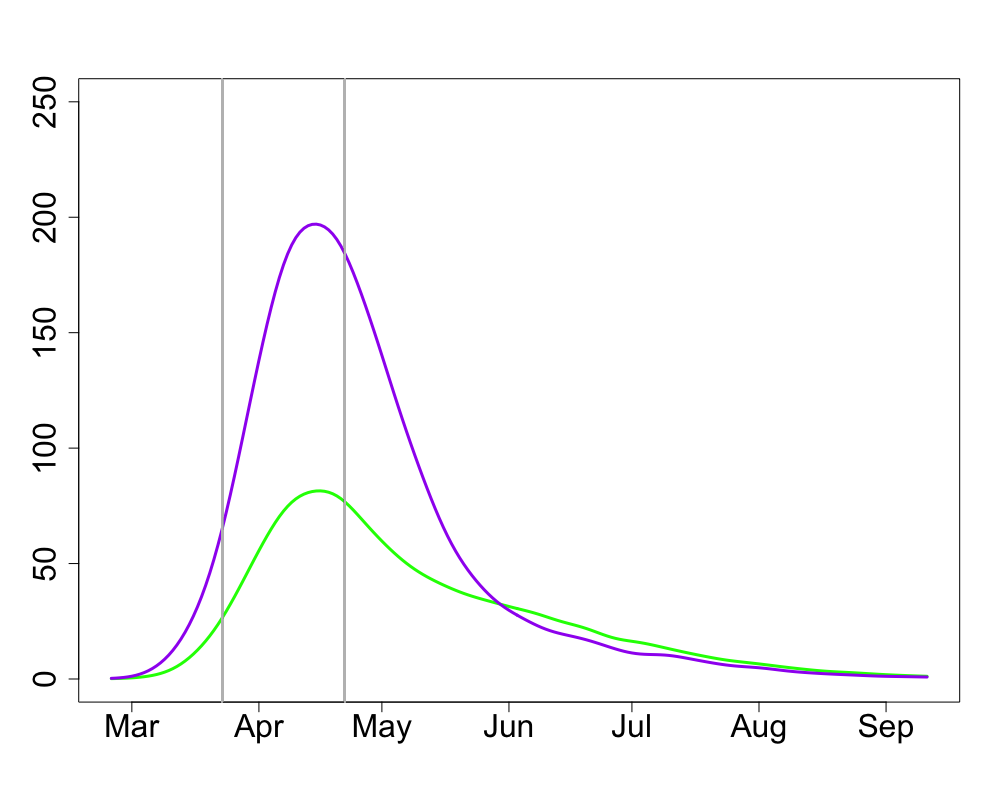}
\caption{Expected deaths at time $t$ at group~$G$ and the rest of population. 
}
\label{fig6}
\end{minipage}
\end{figure}

\section{Sensitivity to main parameters of the model and exit strategy}
\label{sens}
We have selected  Figures~\ref{fig5} and \ref{fig6} as the main figures. The next pairs of figures show sensitivity of the model with respect to different parameters in the model. All these pairs of figures are similar  Figures~\ref{fig5} and \ref{fig6} except one of the parameters is changing.
Moreover, all the curves from Figures~\ref{fig5} and \ref{fig6}  are reproduced in the pairs of figures below in the same colours.

The colour scheme in all figures below with even numbers is:
\begin{itemize}
  \item Solid purple: $I_G(t)/n$, the proportion of infected at time $t$ from group $G$;  main set of parameters;
  \item Solid green: $I(t)/N$, the proportion of infected at time $t$ from the population;  main set of parameters;
  \item Solid (and perhaps, dashed) red: $I_G(t)/n$ for the alternative value (values) of the chosen parameter;
  \item Solid (and perhaps, dashed) blue: $I(t)/N$ for the alternative value (values) of  the chosen parameter.
\end{itemize}

The colour scheme in all figures below with odd numbers is:
\begin{itemize}
  \item Solid purple: $D_G(t)$, the expected number of death at time $t$ at  group $G$;  main set of parameters;
  \item Solid green: $D_{other}(t)$,  the expected number of death at time $t$ for the rest of population;  main set of parameters;
  \item Solid (and perhaps, dashed) red: $D_G(t)$  for the alternative value (values) of the chosen parameter;
  \item Solid (and perhaps, dashed) blue: $D_{other}(t)$ for the alternative value (values) of  the chosen parameter.
\end{itemize}

\subsection{Sensitivity to $t_2$, the time of lifting the lock-down}

\label{sec:t2}

In Figures~\ref{fig9} and \ref{fig10}, we have moved the time $t_2$, the time of lifting the lock-down restrictions, 9 days forward so that the lock-down period is 21 days rather than 30 days as in the main scenario. The  solid    red and blue colours in Figures~\ref{fig9} and \ref{fig10} correspond to $t_2={\rm April\; 13}$ and the middle grey vertical line marks ${\rm April\; 13}$.

\begin{figure}[h]
\centering
\begin{minipage}{.45\textwidth}
  \centering
  \includegraphics[width=1\textwidth]{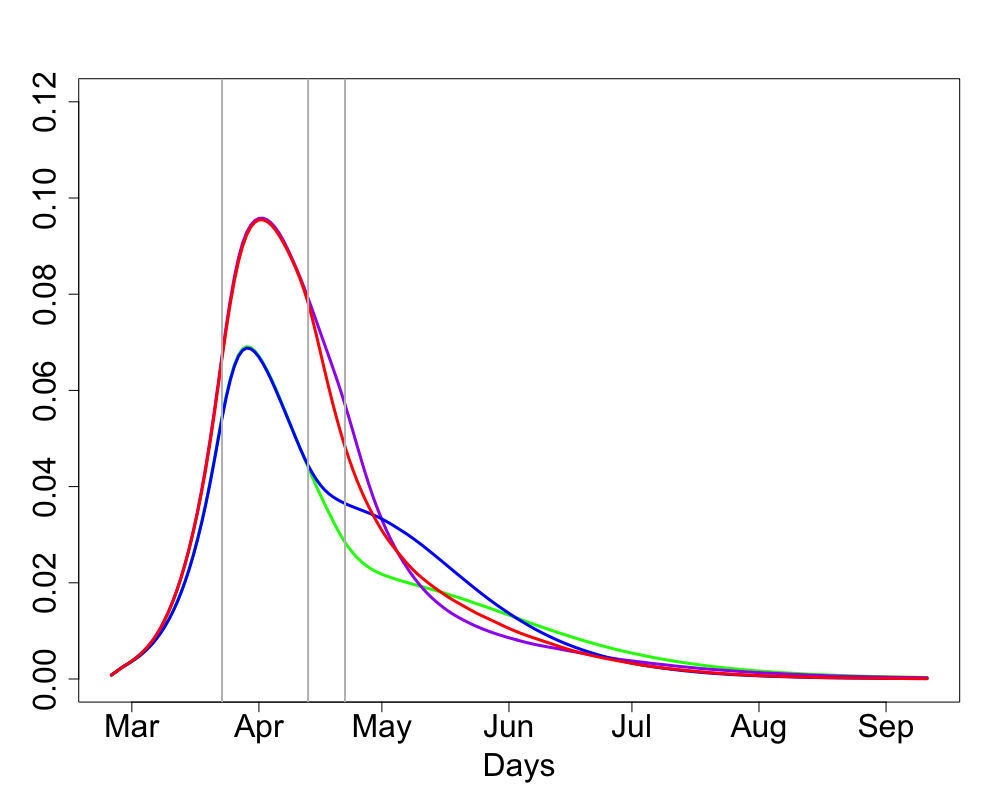}
\caption{Proportions of people infected \\ at time $t$; $t_2{}={\rm April\; 13, 22}$}
\label{fig9}
\end{minipage}%
\begin{minipage}{.45\textwidth}
  \centering
\includegraphics[width=1\textwidth]{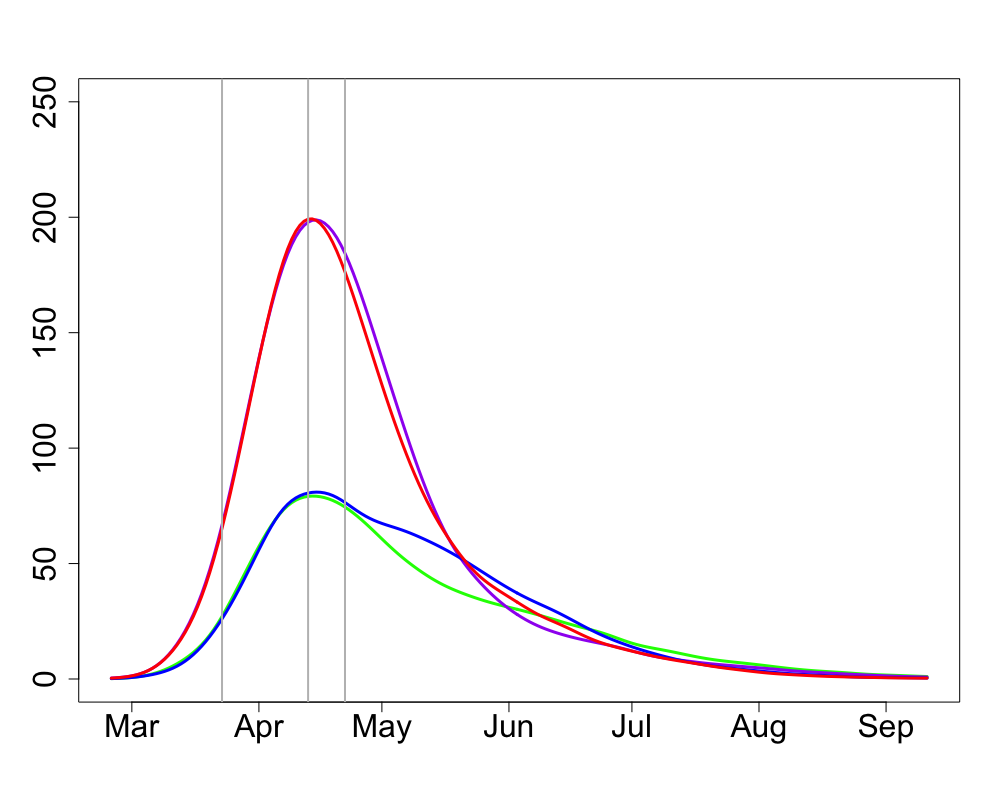}
\caption{Expected deaths at time $t$ at group $G$ and the rest of population;  $t_2{}={\rm April\; 13, 22}$}
\label{fig10}
\end{minipage}
\end{figure}

Consider first Figure~\ref{fig9} showing proportions of infected people and reflecting the demand for hospital beds. Since in the scenario
$t_2={\rm April\; 13}$
 we start isolating people from $G$ earlier than for $t_2={\rm April\; 22}$, the number of infected from group $G$ (red solid line) is slightly lower  during the end of April -- beginning of May for $t_2={\rm April\; 13}$ (purple solid line). The total number of infections from the rest of population is clearly slightly higher for $t_2={\rm April\; 13}$ (blue line) than for $t_2={\rm April\; 22}$. Taking into account the fact that infected people from $G$ would on average require more hospital beds than the rest of population, the expected number of hospital beds required  for the whole population stays approximately the same during the whole epidemic which has to be almost over in July.

 Figure~\ref{fig10}, showing expected deaths numbers for both scenarios with $t_2={\rm April\; 13}$ and $t_2={\rm April\; 22}$, show similar patterns concluding that there is very little gain in keeping the lock-down, especially taking into account huge economic loss caused by every extra day of the lock-down \cite{ons1}.
 Overall, the total expected death is higher but the difference can be considered as very small.
Expected deaths tolls for $t_2={\rm April\; 22}$ and $t_2={\rm April\; 13}$ are   15(5.2+9.8)K and  15.3(5.6+9.7)K, respectively.

{\bf Conclusion.} {\it
The lock-down started on March 23 has slowed the speed of COVID-19 epidemic. However, there is very little gain, in terms of the projected hospital bed occupancy and expected numbers of death, of continuing the lock-down beyond
${ April\; 13}$.
}

\subsection{Sensitivity to $c$, the degree of separation of people from group $G$}
\label{sec:c}

In Figures~\ref{fig11} and \ref{fig12},
 solid   and dashed  line styles (for red/blue colours) correspond to $c=1$ and $c=0.5$ respectively. By increasing $c$ we significantly increase the death toll in the group $G$.
 It is clear that the value of $c$ measures the degree of isolation of people from $G$ and has negligible effect on the rest  of population; this is clearly seen in both figures. On the other hand, the value of $c$ has very significant effect on the people from $G$.

\begin{figure}[h]
\centering
\begin{minipage}{.45\textwidth}
  \centering
  \includegraphics[width=1\textwidth]{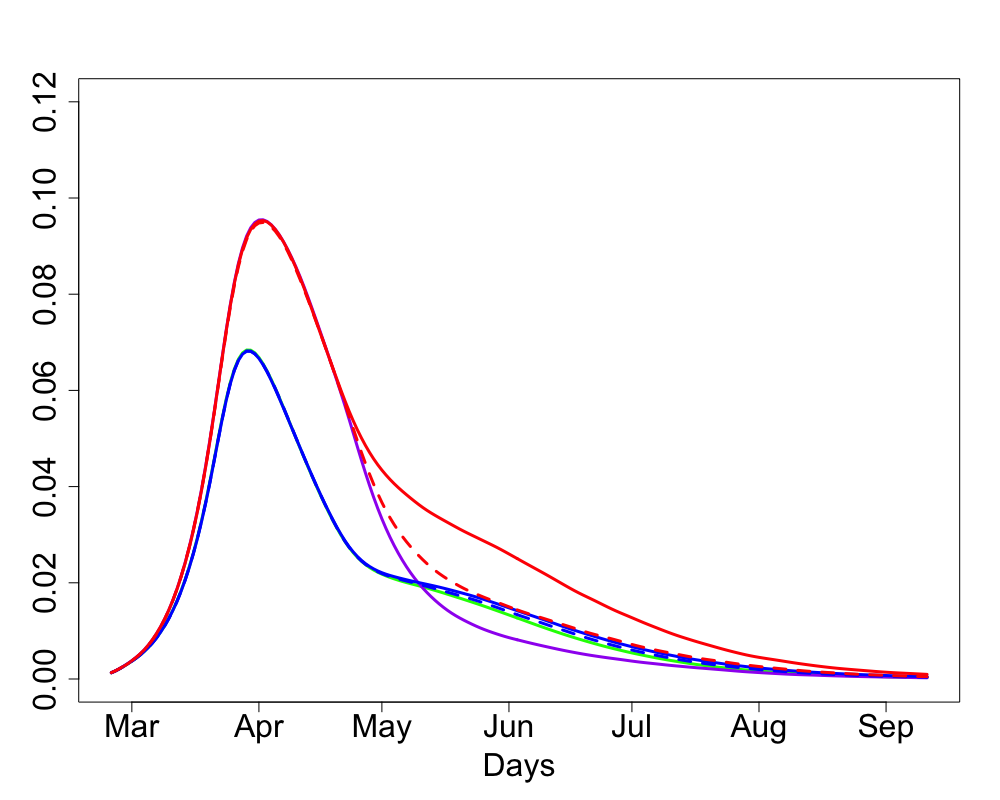}
\caption{Proportions of people infected \\ at time $t$; $c=0.25,0.5,1$}
\label{fig11}
\end{minipage}%
\begin{minipage}{.45\textwidth}
  \centering
\includegraphics[width=1\textwidth]{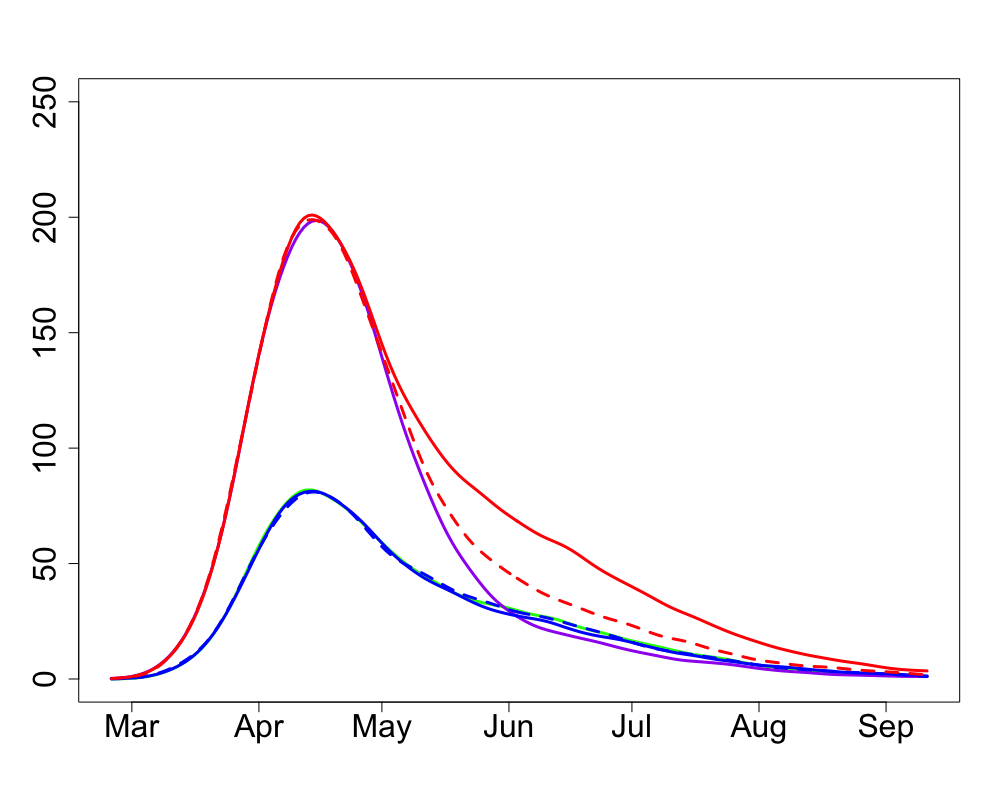}
\caption{Expected deaths at time $t$ at group $G$ and the rest of population; $c=0.25,0.5,1$}
\label{fig12}
\end{minipage}
\end{figure}

 Consider first Figure~\ref{fig11} giving proportions of infected people. One can clearly see that the purple line (group $G$, $c=0.25$) is much lower that the solid red line (group $G$, $c=1$) for all the time until the end of the epidemic, from May until August. In June, in particular,
 the number of infected people from $G$ with $c=0.25$ is  about 40\% of the number of infected people from $G$ with $c=1$.

 Curves on Figure~\ref{fig12}, showing expected deaths at time $t$ in group $G$ and the rest of population, follow the same patterns (with about 2-week time shift) as the related curves of Figure~\ref{fig11} displaying the number of infected. In June and July we should expect
 only about 40\%
 of the number of deaths in
  the scenario with  isolation of people from $G$  $(c=0.25$) relative to the  scenario with no special isolation for people from $G$ ($c=1$).

This is reflected in the expected deaths tolls for entire period of epidemic. Indeed, expected deaths tolls for $c=1$ and 0.5 are   18.2(5.2+13)K and  16.2(5.2+11)K, respectively; compare this with  15(5.2+9.8)K for $c=0.25$.

 {\bf Conclusion.} {\it The effect of the degree of isolation of people from group $G$  in the aftermath of the lock-down is very significant.
}

\subsection{Sensitivity to $R_0$}

In view of recent data, the value of $R_0$ is likely to be lower than $2.5$, especially in small towns and rural areas. As Figures~\ref{fig7} and \ref{fig8} demonstrate, in such places the epidemic will be significantly milder. The overall dynamics of the epidemic would not change much though.

In Figures~\ref{fig7} and \ref{fig8},
 solid   and dashed  line styles (for blue/red colours) correspond to $R_0=2.3$ and $R_0=2.7$ respectively.
Expected deaths tolls for $R_0=2.3, 2.5 $ and $R_0=2.7$ are  14.5(5+9.5)K , 15(5.2+9.8)K and  15.5(5.3+10.2)K, respectively.

\begin{figure}[h]
\centering
\begin{minipage}{.45\textwidth}
  \centering
  \includegraphics[width=1\textwidth]{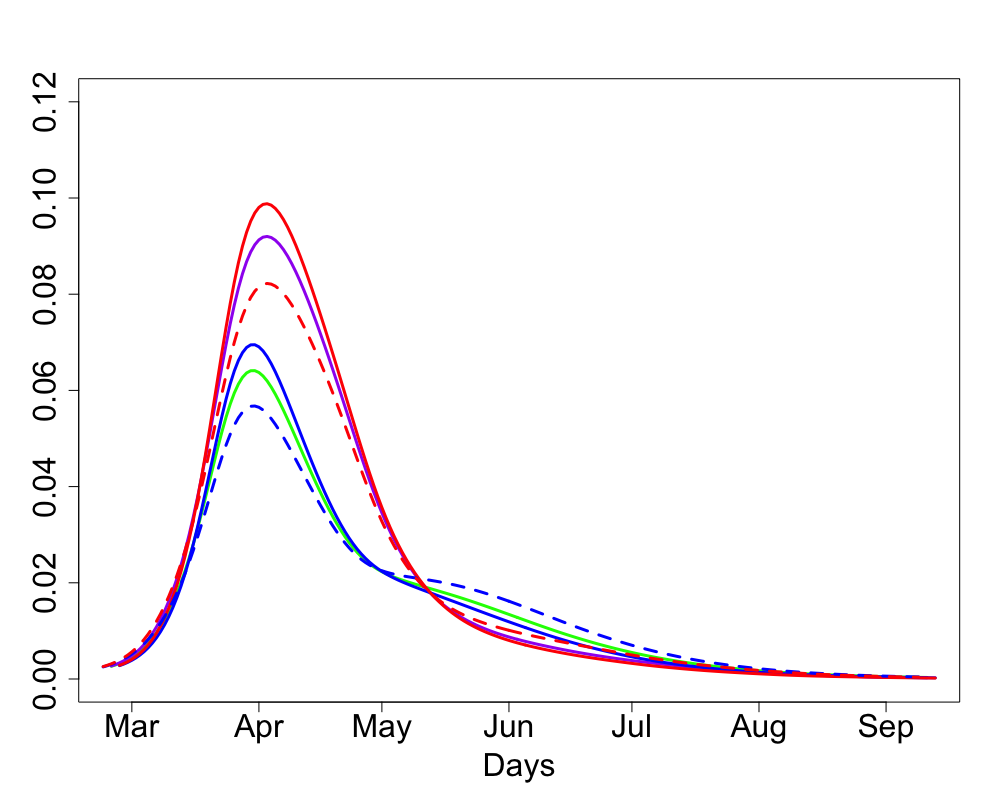}
\caption{Proportions of people infected \\ at time $t$; $R_0=2.3,2.5,2.7$}
\label{fig7}
\end{minipage}%
\begin{minipage}{.45\textwidth}
  \centering
\includegraphics[width=1\textwidth]{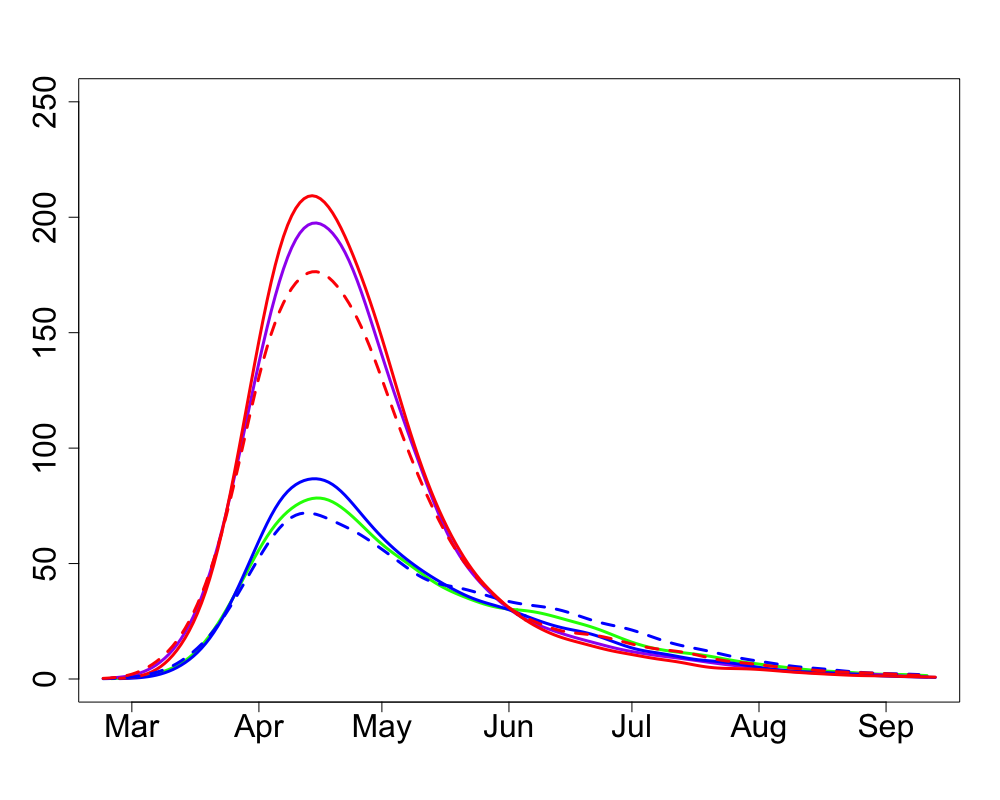}
\caption{Expected deaths at time $t$ at group \\$G$ and the rest of population; $R_0=2.3,2.5,2.7$}
\label{fig8}
\end{minipage}
\end{figure}

\subsection{Sensitivity to $x$, the proportion of infected at the start of the lock-down }
\label{sec:x}

In Figures~\ref{fig21} and \ref{fig22}, we use $x=0.8$ and $x=0.9$.
Expected deaths toll for $x=0.8$ is 17.3(5.3+12)K. This is  higher than 15(5.2+9.8)K for $x=0.9$.
The fact that the difference is significant is  related to the larger number of death for $x=0.8$ in the initial period of the epidemic.
In Figure~\ref{fig21}, red and blue display the number of infected at time $t$ for $x=0.8$ for people from group $G$ and the rest of population, respectively. Similarly, in Figure~\ref{fig22} the red and blue colours show the expected numbers of deaths in these two groups.  Figures~\ref{fig21} and \ref{fig22} show that the timing of the lock-down has serious effect on the development of the epidemic. Late call for a lock-down (when $x=0.8$) helps to slow down the epidemic and guarantees its fast smooth decrease but does not safe as many people as, for example, the call with $x=0.9$.

\begin{figure}[h]
\centering
\begin{minipage}{.45\textwidth}
  \centering
  \includegraphics[width=1\textwidth]{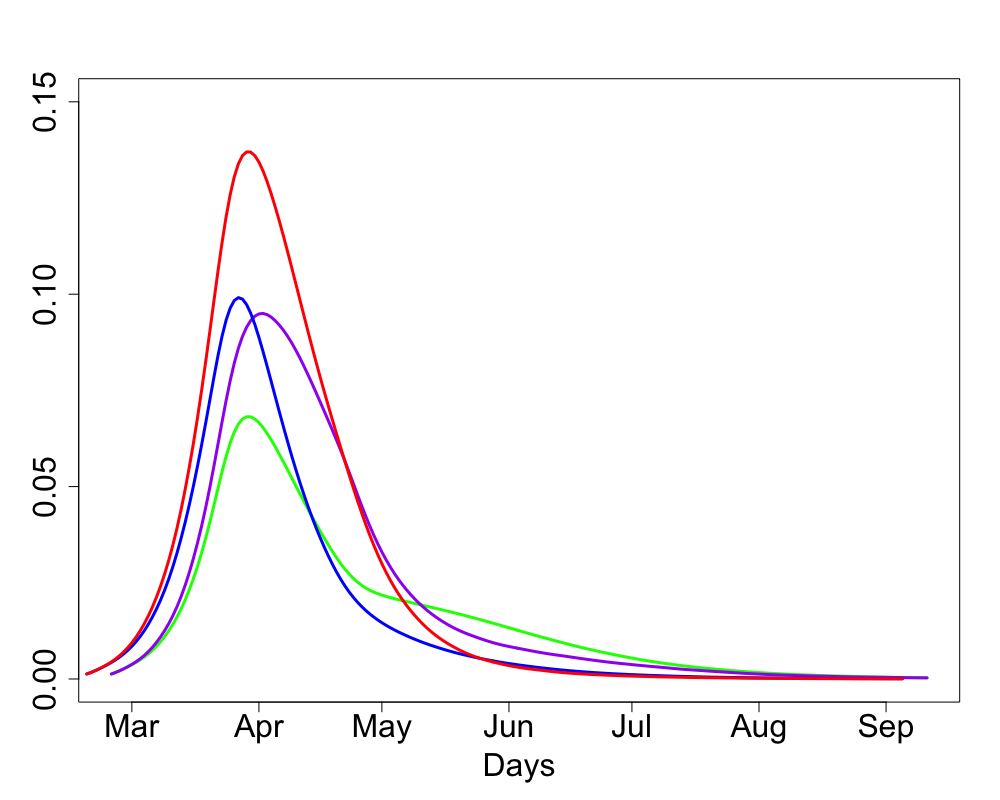}
\caption{Proportions of people infected \\ at time $t$; $x=0.8, 0.9$}
\label{fig21}
\end{minipage}%
\begin{minipage}{.45\textwidth}
  \centering
\includegraphics[width=1\textwidth]{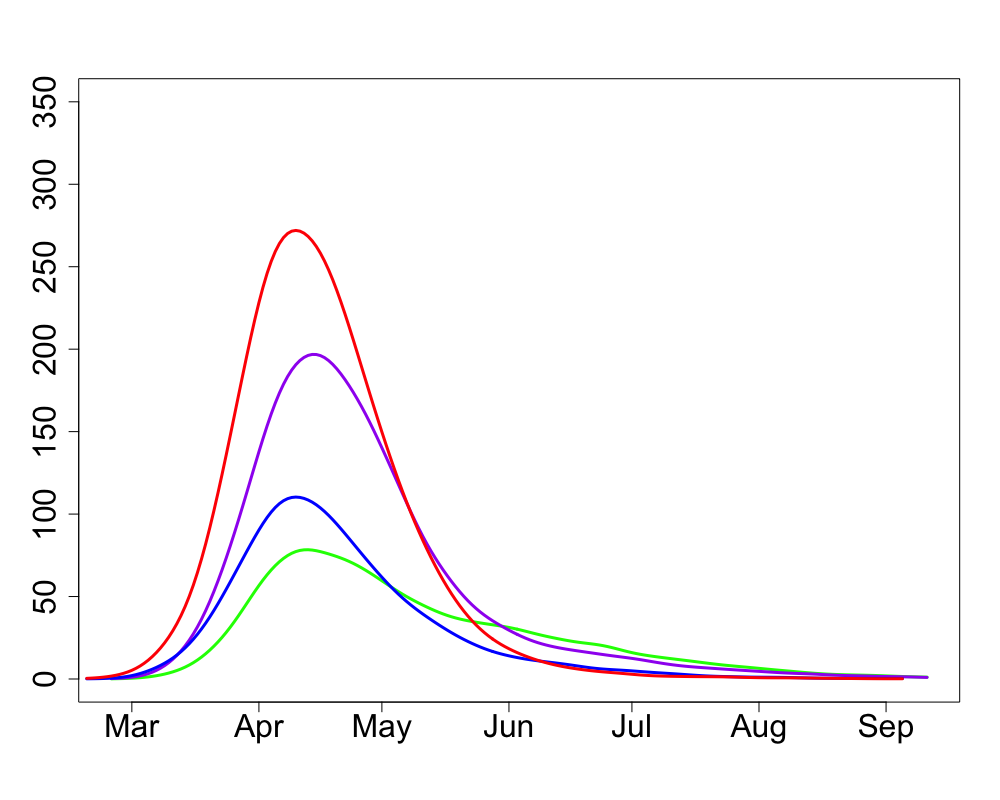}
\caption{Expected deaths at time $t$ at group $G$ and the rest of population; $x=0.8, 0.9$}
\label{fig22}
\end{minipage}
\end{figure}

Note that in both cases, when $x=0.8$ and $x=0.9$, the second wave of the epidemic is not expected as by the time of lifting the lock-down, a large percentage of the population (about 25\% in case $x=0.9$ and almost 40\% in case $x=0.8$) is either infected or immune  and the `herd immunity' would follow shortly. In a certain sense, the decision of making a lock-down at around $x=0.9$ seem to be a very sensible decision to make (considering economic costs of each day of a lock-down) as this saves many lives and guarantees smooth dynamics of the epidemic with no second wave.

Let us now consider  more informative scenarios when $x=0.95$ and $x=0.97$; that is, when the lock-down is made  early, see
Figures~\ref{fig23},\ref{fig24} for $x=0.95, 0.9$ and Figures~\ref{fig25},\ref{fig26} for $x=0.97, 0.9$. With an early lock-down, there is a  clear gain in hospital bed occupancy and expected number of death at the first stage of the epidemic. However, the second wave of the epidemic
should be expected in both cases $x=0.95$ and $x=0.97$  with a peak at around 2 months after the first one, and the second peak could be higher than the first one. This can be explained by observing  that, even after 2 months of an epidemic with large $x$, even with lock-down and strong isolation of the group $G$ and relatively small reproductive number (recall $R_2=2$), there is still a very large proportion of non-immune people 
available for the virus;  a large part of these people  is going to be  infected  even with smaller reproductive number. This prolongs the epidemics. Expected deaths toll for $x=0.95$ is 13.8(5.5+8.3)K and for $x=0.97$ it is 13.4(5.7+7.7)K.
These numbers are  naturally lower than 15(5.2+9.8)K for $x=0.9$ but the difference is not significant.

Figures~\ref{fig23}--\ref{fig26} and the  discussion above imply the following conclusion.

{\bf Conclusion.} 
{\it A lock-down at an early stage of an epidemic (unless it is a very strict one like in Wuhan) is not a sensible decision in view
of the economic consequences of the lock-down and the measures required for all the remaining (much longer) period of the epidemic.
Moreover, in the case of an early lock-down, the second wave of the epidemic should be expected 
with a peak at around 2 months after the first one.}

%

%
%

\begin{figure}[h]
\centering
\begin{minipage}{.45\textwidth}
  \centering
  \includegraphics[width=1\textwidth]{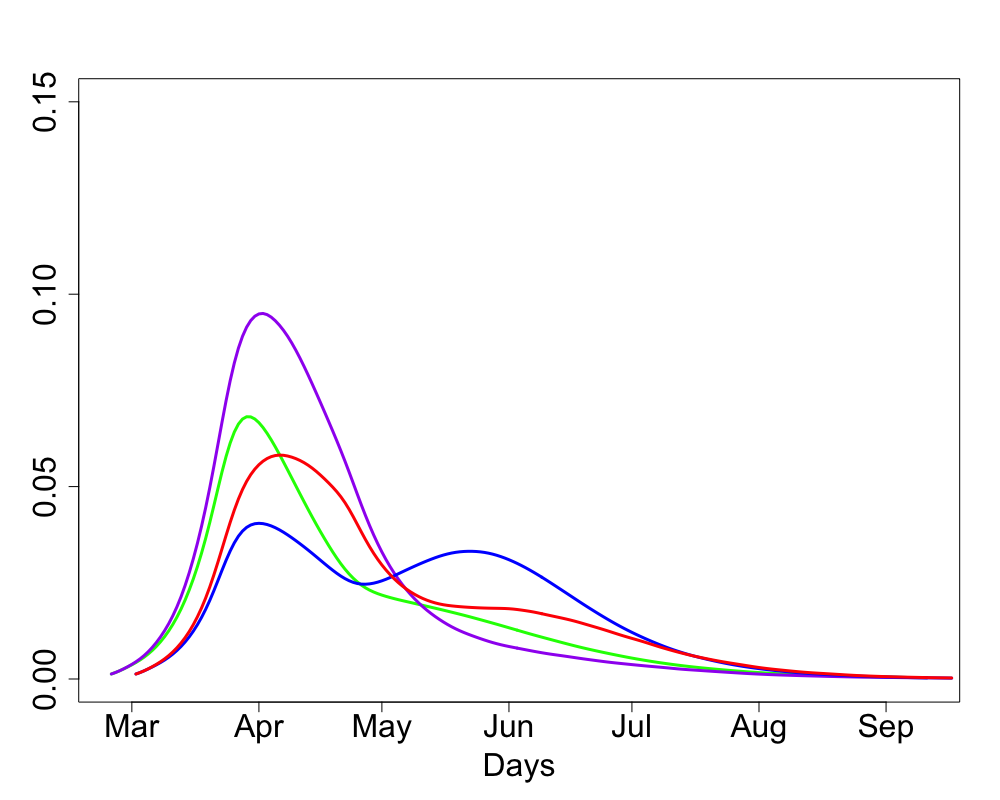}
\caption{Proportions of people infected \\ at time $t$; $x=0.9, 0.95$}
\label{fig23}
\end{minipage}%
\begin{minipage}{.45\textwidth}
  \centering
\includegraphics[width=1\textwidth]{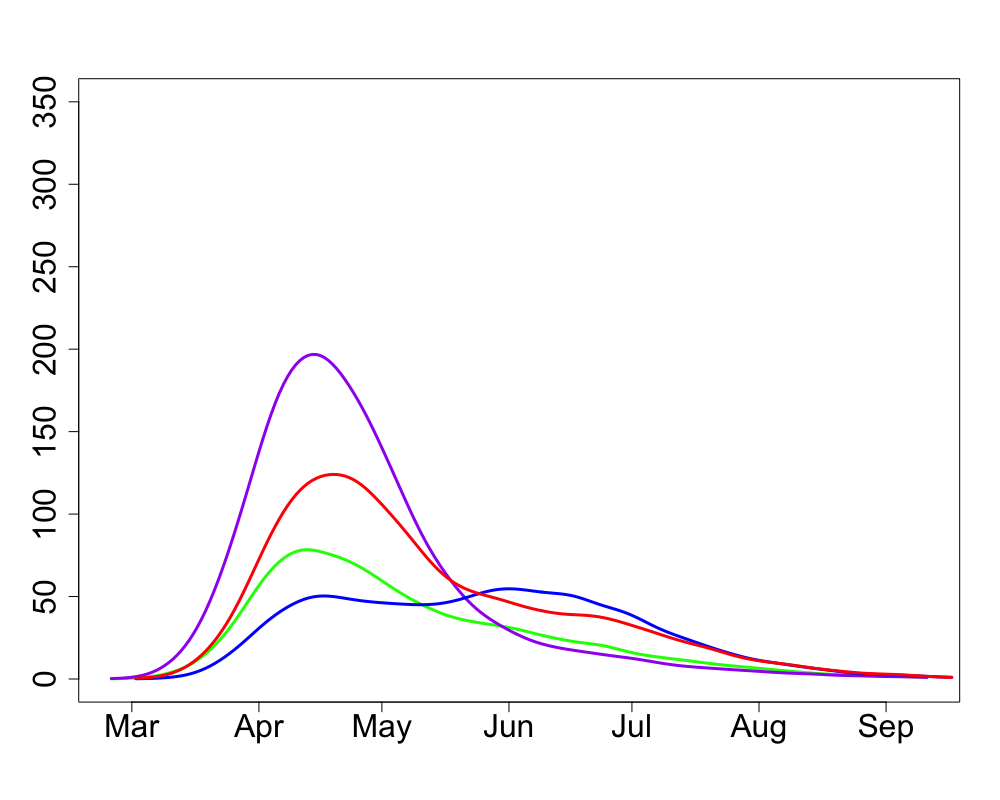}
\caption{Expected deaths at time $t$ at group $G$ and the rest of population; $x=0.9, 0.95$}
\label{fig24}
\end{minipage}
\end{figure}

In Figures~\ref{fig25} and \ref{fig26}, we use $x=0.97$ and $x=0.9$.
Expected deaths toll for $x=0.97$ is 13.8(5.7+8.3)K. This is naturally lower than 15(5.2+9.8)K for $x=0.9$.
This is related to the fact that we isolate people from group $G$ much earlier.

A disadvantage of this scenario is the second wave of epidemic with a peak at around 2 months after the first one.

\begin{figure}[h]
\centering
\begin{minipage}{.45\textwidth}
  \centering
  \includegraphics[width=1\textwidth]{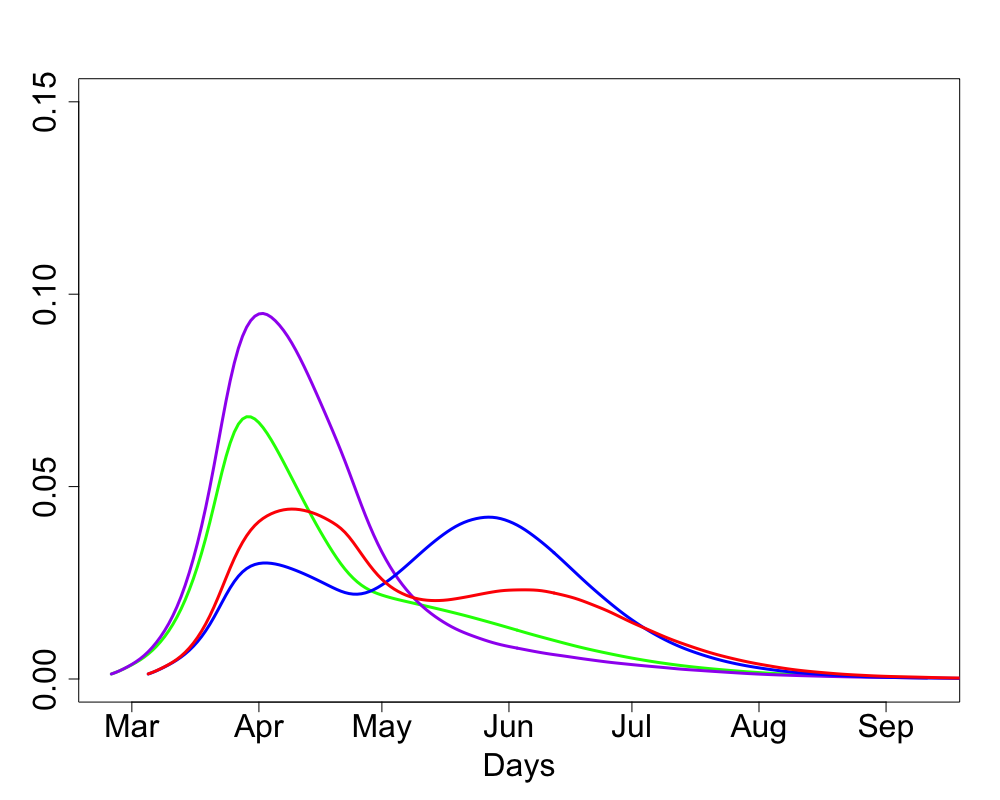}
\caption{Proportions of people infected \\ at time $t$; $x=0.9, 0.97$}
\label{fig25}
\end{minipage}%
\begin{minipage}{.45\textwidth}
  \centering
\includegraphics[width=1\textwidth]{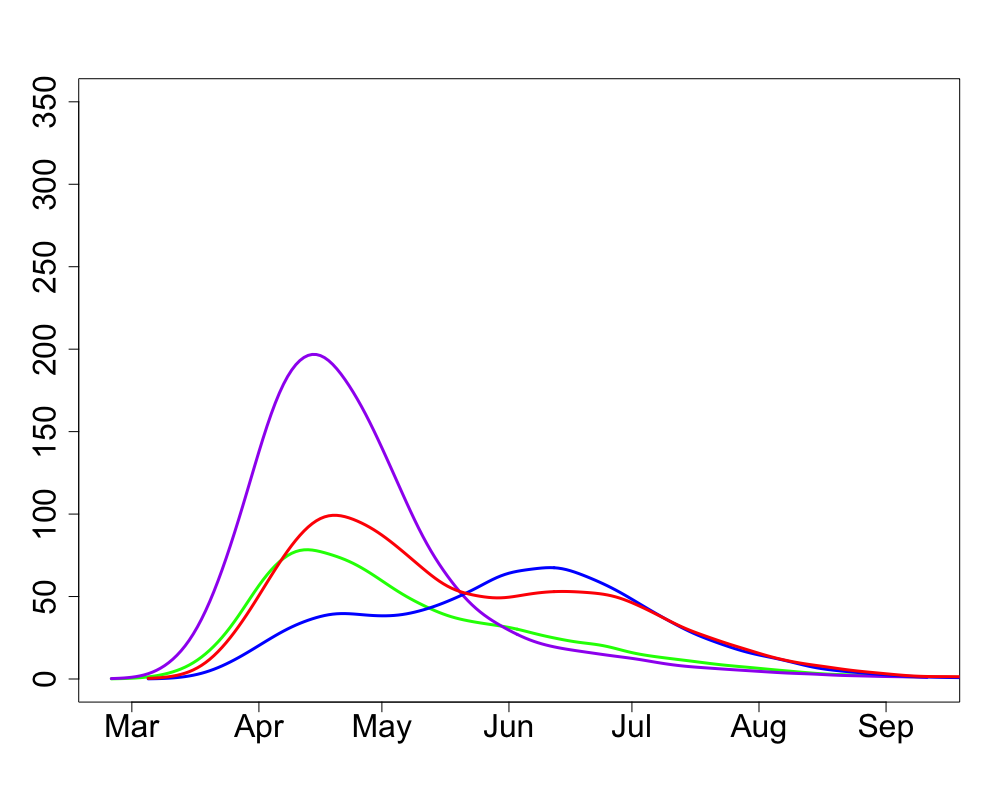}
\caption{Expected deaths at time $t$ at group $G$ and the rest of population; $x=0.9, 0.97$}
\label{fig26}
\end{minipage}
\end{figure}

\subsection{Sensitivity to $R_2$, the reproductive number after lifting the lock-down}

\label{R2}

In Figures~\ref{fig19} and \ref{fig20}, we use $R_2=2.0$ and $2.5$. Increase in $R_2$ would lead to a significant increase in  the number of severe cases and  expected death  numbers.

Expected deaths toll for $R_2=2.5 $ is 16.9(5.9+11)K. This is  significantly higher than 15(5.2+9.8)K for $R_2=2$.
This implies that the public should carry on some level of isolation in the next 2-3 months.

{\bf Conclusion.} {\it Increase in the reproductive number after lifting the lock-down would inevitably imply  a significant increase in  the number of severe cases and  expected death  numbers.}

\begin{figure}[h]
\centering
\begin{minipage}{.45\textwidth}
  \centering
  \includegraphics[width=1\textwidth]{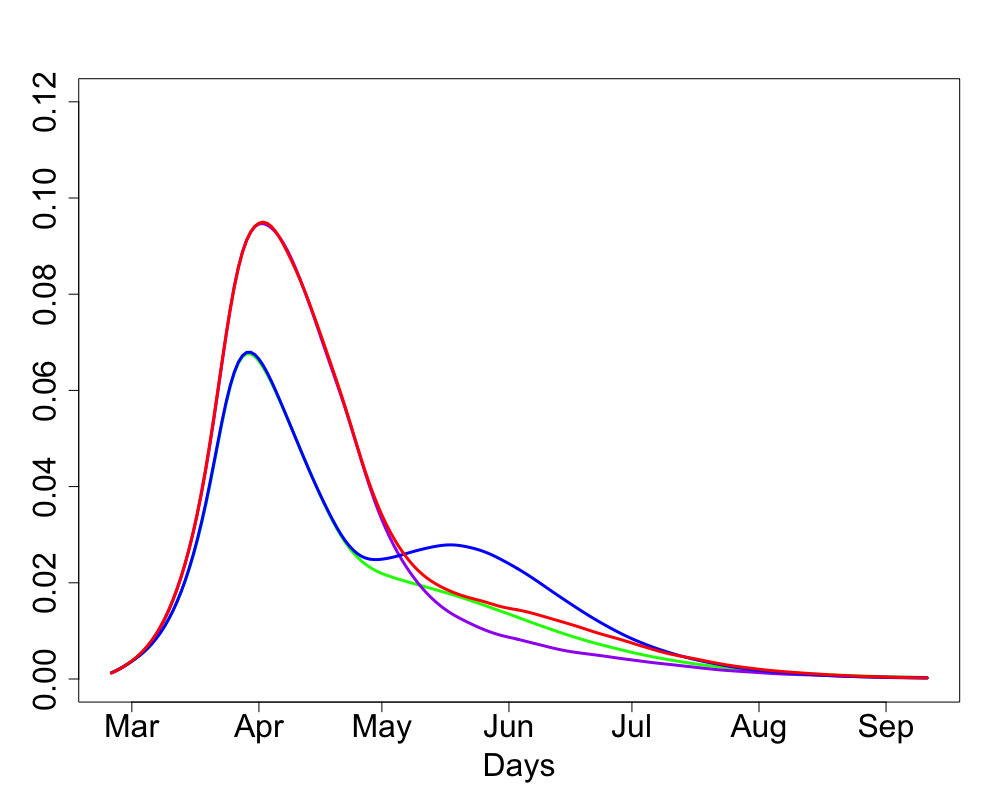}
\caption{Proportions of people infected \\ at time $t$; $R_2=2, 2.5$}
\label{fig19}
\end{minipage}%
\begin{minipage}{.45\textwidth}
  \centering
\includegraphics[width=1\textwidth]{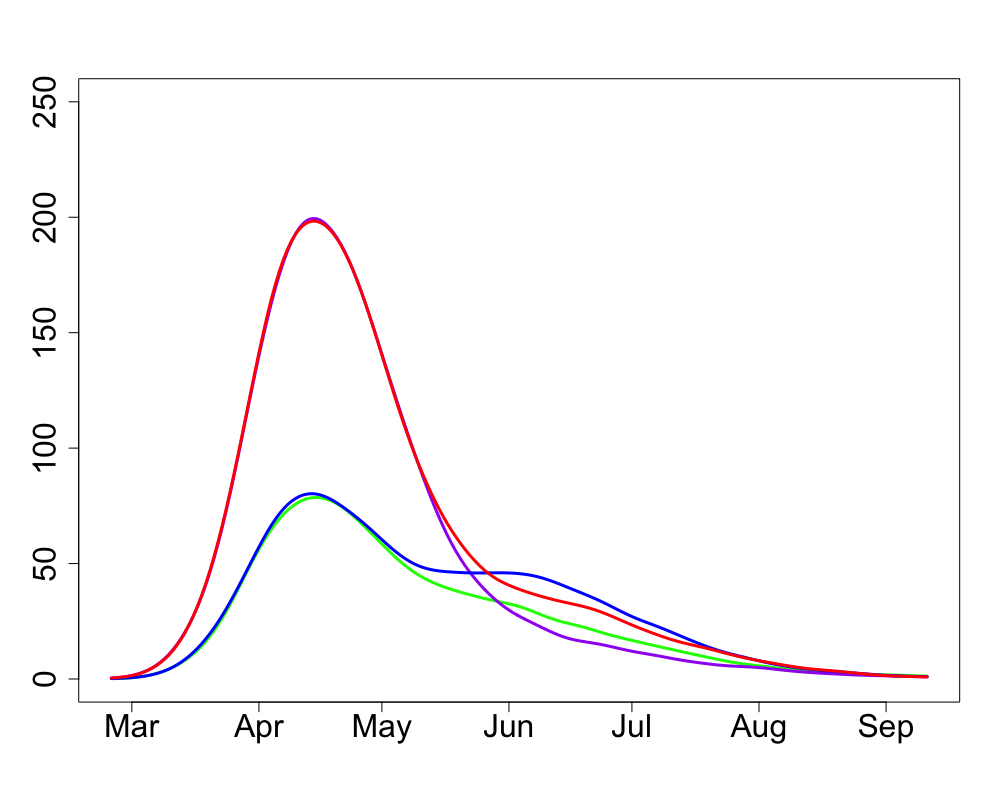}
\caption{Expected deaths at time $t$ at group $G$ and the rest of population; $R_2=2, 2.5$}
\label{fig20}
\end{minipage}
\end{figure}

\subsection{Sensitivity to $k_M$ and $k_S$, the shape parameter of the Erlang distributions for mild and severe cases
}

In Figures~\ref{fig13} and \ref{fig14}, we use $k_M=1,3$. The value $k_M=3$ is default while the value $k_M=1$ defines the exponential distribution for the period of infection in the case of mild disease and is equivalent to the corresponding assumption in SIR models. As the variance of the exponential distribution is larger than of the Erlang with $k_M=3$ (given the same means), the epidemic with $k_M=1$ runs longer and smoother.

\begin{figure}[h]
\centering
\begin{minipage}{.45\textwidth}
  \centering
  \includegraphics[width=1\textwidth]{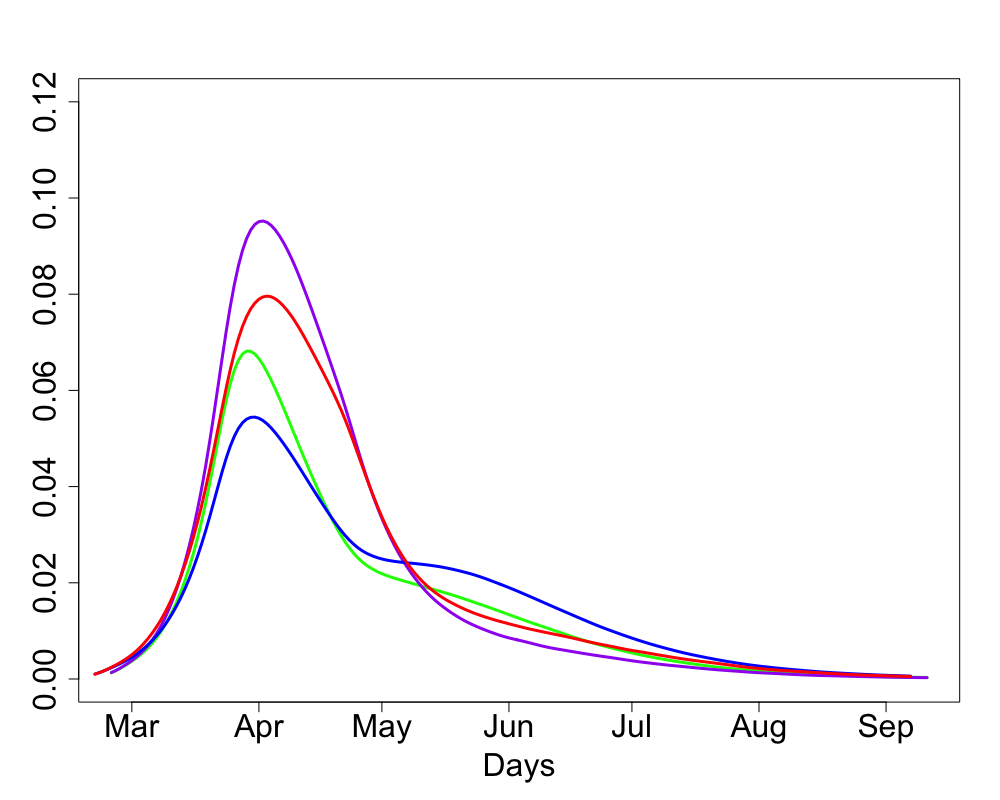}
\caption{Proportions of people infected \\ at time $t$; $k_M=1,3$}
\label{fig13}
\end{minipage}%
\begin{minipage}{.45\textwidth}
  \centering
\includegraphics[width=1\textwidth]{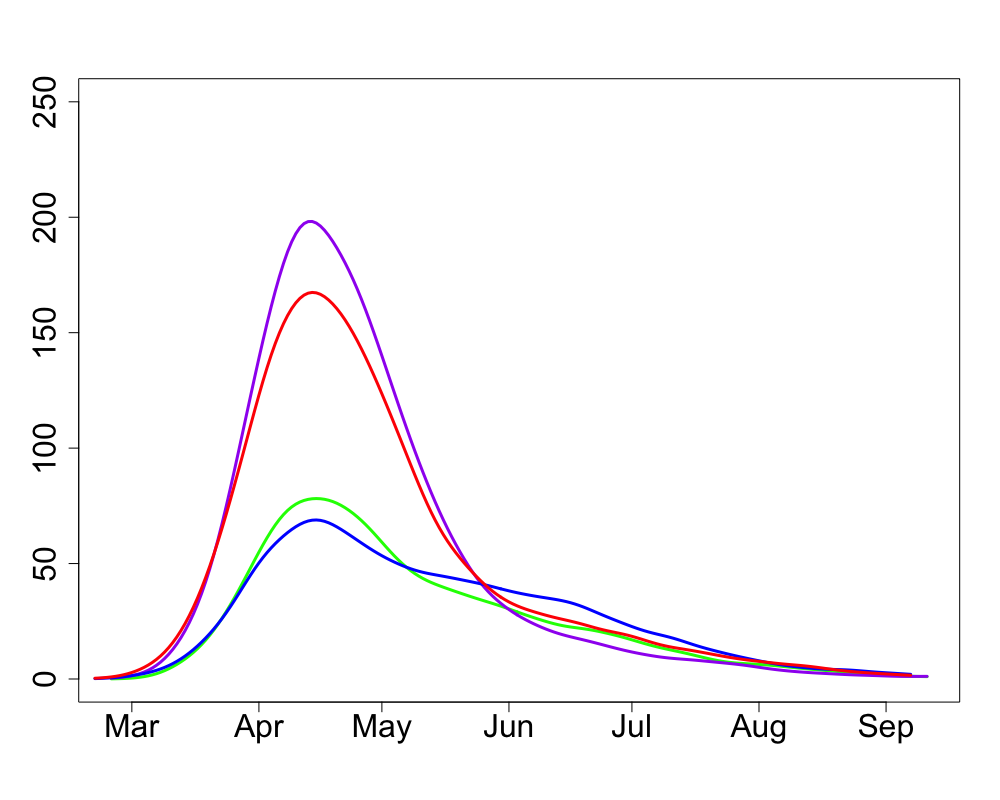}
\caption{Expected deaths at time $t$ at group $G$ and the rest of population; $k_M=1,3$}
\label{fig14}
\end{minipage}
\end{figure}


In Figures~\ref{fig15} and \ref{fig16}, we use $k_S=1,3$. Parameter $k_S$ is less sensitive than $k_M$ (as there are less severe cases than the mild ones)  but it is also is.

\begin{figure}[h]
\centering
\begin{minipage}{.45\textwidth}
  \centering
  \includegraphics[width=1\textwidth]{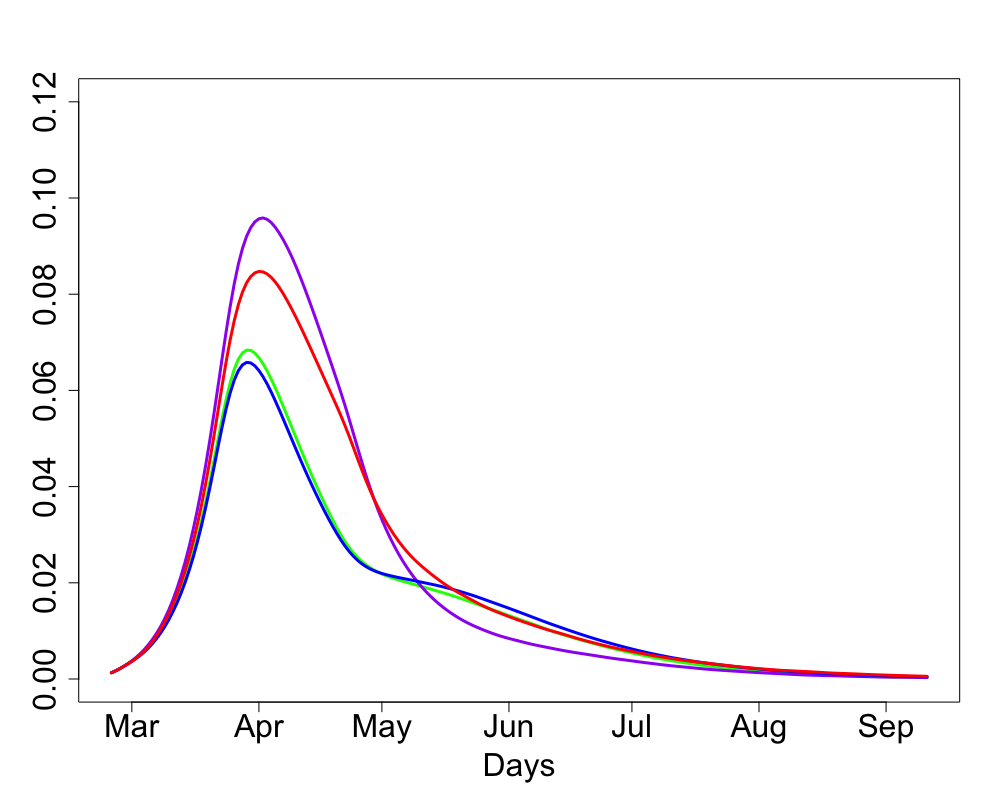}
\caption{Proportions of people infected \\ at time $t$; $k_S=1,3$}
\label{fig15}
\end{minipage}%
\begin{minipage}{.45\textwidth}
  \centering
\includegraphics[width=1\textwidth]{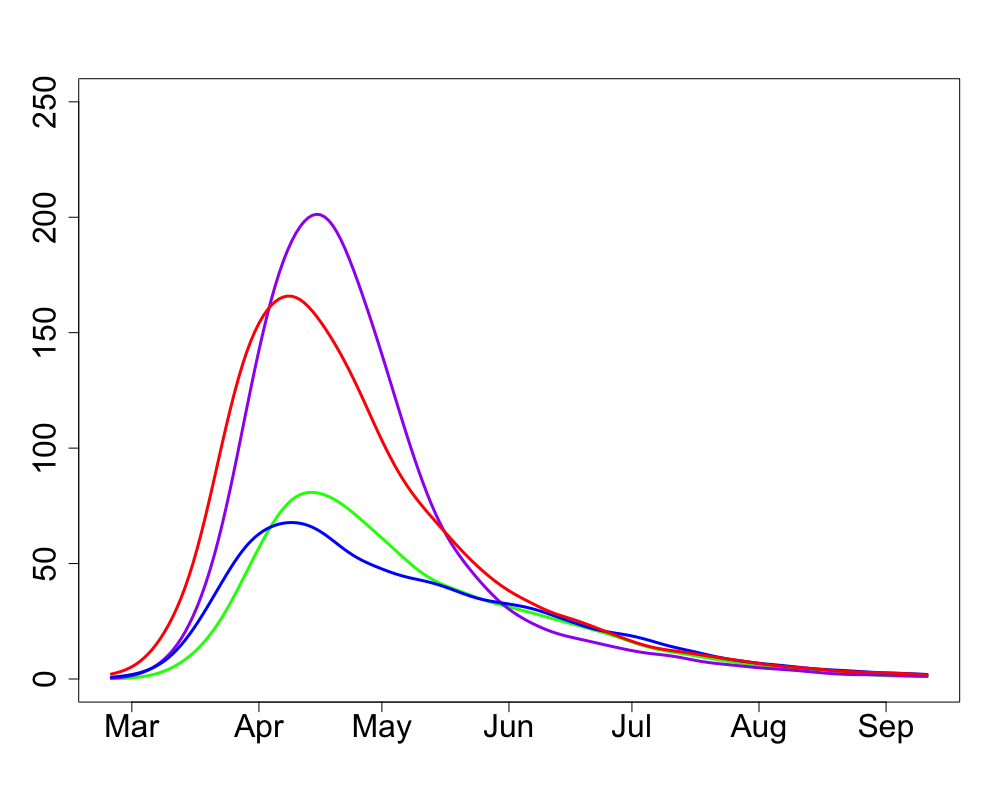}
\caption{Expected deaths at time $t$ at group $G$ and the rest of population; $k_S=1,3$}
\label{fig16}
\end{minipage}
\end{figure}

{\bf Summary.}  {\it The parameter $k_M$ is rather important and more information is needed about the distribution time of infectious period.
Parameter $k_S$ is less sensitive than $k_M$   but it is also is.}

\subsection{Sensitivity to $\delta$, the probability of death in severe cases}

In Figures~\ref{fig17} and \ref{fig18}, we use $\delta= 0.2$ and $0.1$. Decrease of $\delta$ increases the number of severe cases 
but does not change the expected death  numbers.

\begin{figure}[h]
\centering
\begin{minipage}{.45\textwidth}
  \centering
  \includegraphics[width=1\textwidth]{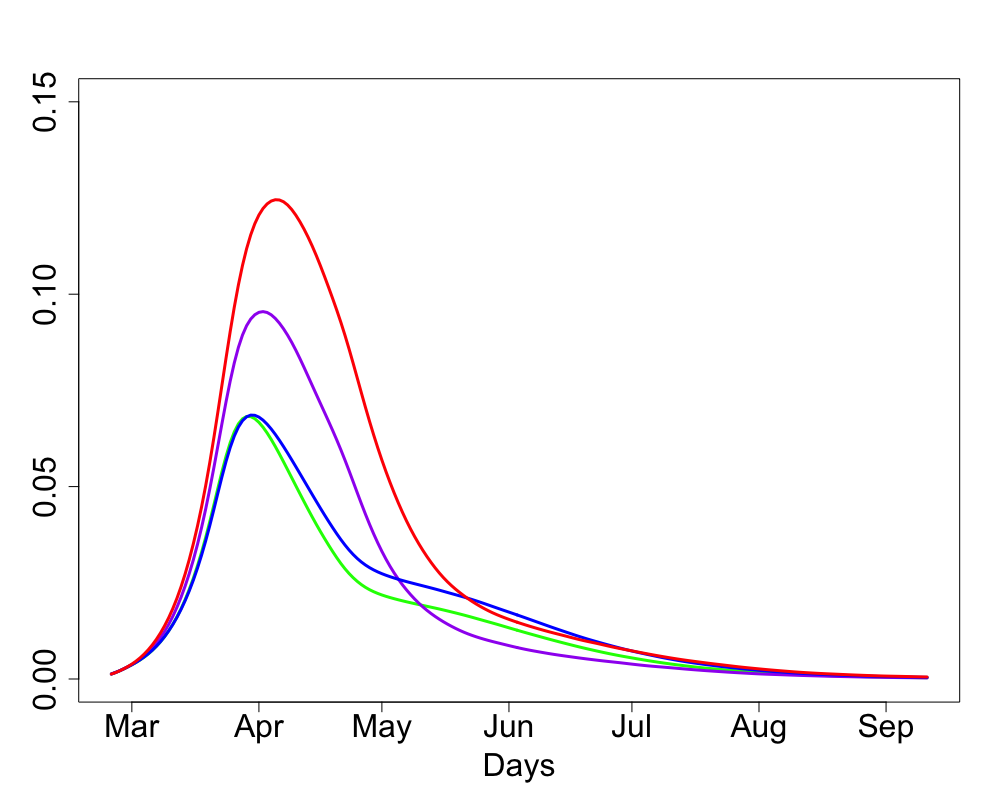}
\caption{Proportions of people infected \\ at time $t$; $\delta= 0.1,0.2$}
\label{fig17}
\end{minipage}%
\begin{minipage}{.45\textwidth}
  \centering
\includegraphics[width=1\textwidth]{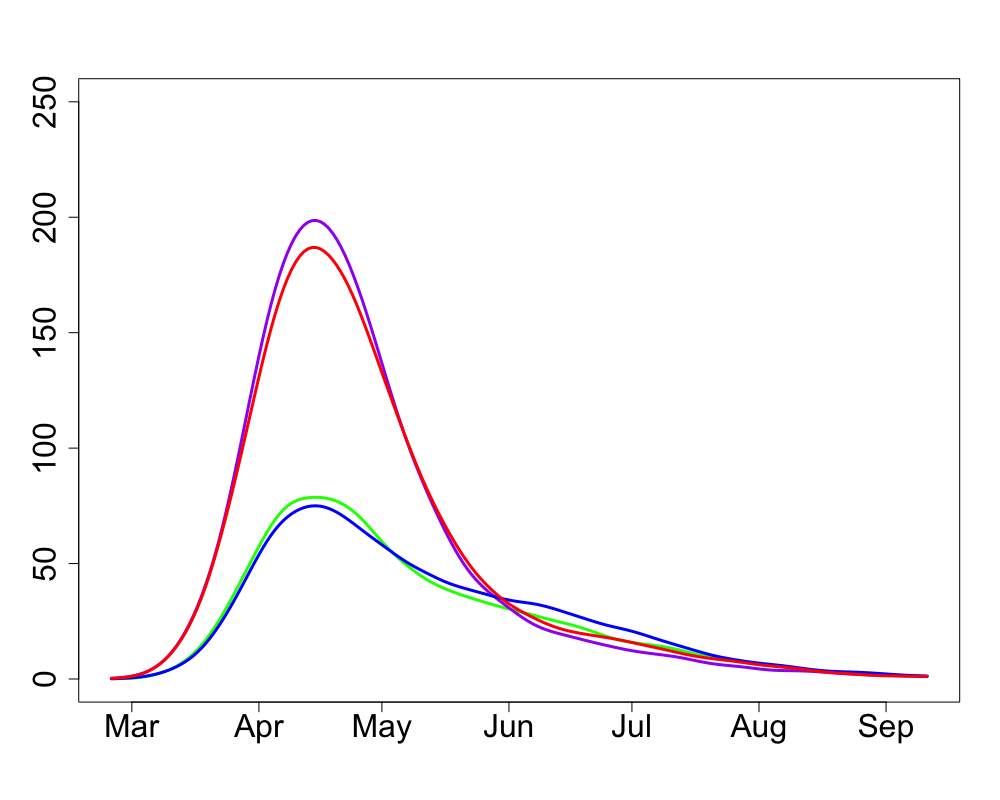}
\caption{Expected deaths at time $t$ at group $G$ and the rest of population; $\delta= 0.1,0.2$}
\label{fig18}
\end{minipage}
\end{figure}



\bibliographystyle{unsrt}
\bibliography{Coronavirus}

\end{document}